\documentstyle[]{article}

\title{
       Approximate Lie Group Analysis of a Model
       Advection Equation on an Unstructured Grid
      }

\author{
         {
          A.M. Latypov} \\
         \mbox{ } \\
         \it  Department of Mathematics and Statistics\\
         \it  and Fluid Dynamics Research Institute,\\
         \it  University of Windsor, 401 Sunset Ave.,\\
         \it  Windsor, Ontario, Canada N9B 3P4\\
         \it  Internet email: latypov@server.uwindsor.ca
       }
\date{
     \mbox{\scriptsize{ } }
     }
\begin{document}
\newtheorem{definition}{Definition}
\newtheorem{theorem}{Theorem}
\newtheorem{lemma}{Lemma}

\bibliographystyle{plain}

\maketitle

\begin{abstract}
\noindent{\bf Purpose:} To investigate the Lie groups of
transformations (symmetries) approximately admitted by a
finite difference discretization of a nonlinear advection equation
on an unstructured mesh.\\
{\bf Method:} A recently developed technique of
``approximate group analysis'' is applied to a differential
approximation (otherwise referred to as an equivalent
differential equation) corresponding to the above finite difference
approximation.\\
{\bf Results and Conclusions:} We determine which groups from
the infinite variety of groups admitted by a nonlinear advection
equation ``survive'' the discretization. The situations arising
for different choices of an arbitrary function (local speed of
propagation) are also studied.
\end{abstract}

\section{Introduction}
Invariance of  differential equations with respect to
one--parameter groups of transformations in the space of
independent and dependent variables carries important
information about the fundamental properties of the physical
system that these equations describe, such as
conservation laws.
Knowledge of the groups admitted by the system of differential
equations may also allow one to reduce the order of the system,
to find important particular solutions, or to produce new solutions
from a single solution which is known \cite{OLVER}.

Preserving the group properties of differential equations
in their discretizations seems to be a natural requirement
which would ensure that the fundamental properties of the
approximated differential equations are preserved in the discrete
models \cite{SHOKIN:1, MAEDA, WINTERNITZ, DORODNITSYN}.
This idea is often followed in the design of approximations
used in CFD by making sure that the scheme preserves
steady--state or lower--dimensional solutions exactly.
Further illustrating this statement, the author of \cite{SHOKIN:1}
refers to the result from \cite{SHOKIN:2} where it was demonstrated
that the undesirable process of self--oscillation observed
in the solutions obtained using Harlow's scheme \cite{HARLOW}
is caused by the non--invariance of the scheme with respect
to the Galileo transformation.
Unfortunately, in general, replacing the system of differential
equations with their finite difference, finite volume or
finite element discretizations introduces the non--invariance
into the discretization. This non--invariance stems from the fact
that performing the discretization requires utilizing a discrete set
of nodes in the space of independent variables, i.e. a computational grid.
Formally, this implies that the group of transformations acting on the
discrete variables needs to leave invariant not only the approximating
discrete system, but also a set of algebraic equations defining
the grid \cite{DORODNITSYN}. Although some of the important groups
of transformations arising in applications do satisfy this property,
generally speaking, the discrete system
loses many of the groups of the original differential equations.

The natural way to avoid this restrictiveness refered to above
is to weaken the requirement of invariancy of the discrete
system. One of the possibilities is to consider the
{\em differential approximation} \cite{SHOKIN:1} or {\em equivalent
hyperbolic differential equation (EHDE)} \cite{KARNEY}  corresponding to
the discrete system. By formally applying a Taylor series
expansion to the approximating discrete system, one can obtain a
modified differential
equation which can be treated as a perturbation of the original
differential equation due to discretization.

The purpose of this work is to extend the above approach
using  a recently developed concept of {\em approximate invariance}
\cite{BAIKOV:1}, combining the methods of group analysis and
perturbation theory.

\section{Method}

\subsection{Nonlinear advection equation and its symmetries}
We consider a nonlinear advection equation
\begin{eqnarray}
     L(x,t,u,u_t,u_x) \equiv u_t + \varphi(u) u_x = 0,
     \label{eq:advection}
\end{eqnarray}
where $\varphi(u)$ is a given arbitrary smooth function,
characterizing the local speed of propagation of small
disturbances.

Let ${\bf z} = (t,x,u)$ be a  point in the space of independent
variables and unknown functions.
Consider the following one--parameter family of
transformations acting on ${\bf z}$:
\begin{eqnarray}
  {\bf z}^\prime = {\bf f}({\bf z},  a), \label{eq:transformation}
\end{eqnarray}
where ${\bf f} = (f_1,f_2,f_3)$ and $a$ is a scalar parameter of this family
of transformations.
We say that a transformation (\ref{eq:transformation}) is a
{\em one--parameter group (o.p.g.)} (or {\em a symmetry})
with respect to the parameter $a$ if (i)
 ${\bf f}({\bf z},a) = {\bf z}$
 for all $\bf z$, if and only if $a=0$, and (ii)
 ${\bf f}({\bf f}({\bf z},a),b) = {\bf f}({\bf z}, a + b)$
  for all  $a$, $b$ and  ${\bf z}$.
It is convenient to associate
each o.p.g.  with its {\em infinitesimal operator}
\begin{eqnarray*}
 X = \mbox{\boldmath $\xi$}^T \partial_{\bf z}
   = \xi^{(t)}(x,t,u) \partial_t + \xi^{(x)}(x,t,u) \partial_x
   + \eta(x,t,u) \partial_u,
\end{eqnarray*}
where $\mbox{\boldmath $\xi$} = (\xi^{(t)}, \xi^{(x)}, \eta)^T =
 \frac{\partial}{\partial a} {\bf f} {\mid}_{a=0}$. Lie theorem
\cite{OLVER} establishes a one--to--one correspondence between
an o.p.g. and its infinitesimal operator.

Once the o.p.g. of transformations (\ref{eq:transformation}) in
the space of independent variables and unknown
functions is specified, one can easily derive the corresponding
transformations of the partial derivatives of unknown functions
with respect to the independent variables. For  equation
(\ref{eq:advection}), this procedure applied to  partial
derivatives of  first order, results in an o.p.g. of transformations
acting in the space of variables $(x,t,u,u_t,u_x)$, which is
called the first {\em prolongation} of the o.p.g
(\ref{eq:transformation}). One of
the conveniences of using the infinitesimal operator to
describe the o.p.g. is that there exist a compact
{\em prolongation formulae} \cite[Theorem 2.36]{OLVER}, which
gives the components of the first prolongation $X^{(1)}$ of
the infinitesimal operator in terms of the components of
the operator $X$. For the case under consideration, these
formulae give:
\begin{eqnarray*}
 X^{(1)} = X + \zeta^{(t)} \partial_{u_t}
                            + \zeta^{(x)} \partial_{u_x},
\end{eqnarray*}
where
\begin{eqnarray*}
  \begin{array}{ll}
   \zeta^{(t)} = D_t(\eta - \xi^{(t)} u_t - \xi^{(x)} u_x)
                    + \xi^{(t)} u_{tt} + \xi^{(x)} u_{tx}, &
   \zeta^{(x)} = D_x(\eta - \xi^{(t)} u_t - \xi^{(x)} u_x)
                    + \xi^{(t)} u_{tx} + \xi^{(x)} u_{xx},
  \end{array}
\end{eqnarray*}
and $D_t = \partial_t + u_t \partial_u + u_{tt} \partial_{u_t}
                                       + u_{tx} \partial_{u_x} + ...$,
    $D_x = \partial_x + u_x \partial_u + u_{tx} \partial_{u_t}
                                       + u_{xx} \partial_{u_x} + ...$,
are the first total derivatives.

The main problem of group analysis of differential equations
is to find all o.p.g. with respect to which the given differential
equation is invariant. The solution to this problem can generally
be found using {\em infinitesimal criterion of invariance}
\cite[Theorem 2.31]{OLVER} which, for the present case,
reduces to the following statement. Equation (\ref{eq:advection})
is invariant with respect to an o.p.g. (\ref{eq:transformation})
if and only if
\begin{eqnarray*}
   X^{(1)} (u_t + \varphi(u) u_x) {\mid}_{u_t = - \varphi(u) u_x} = 0.
\end{eqnarray*}

Direct calculation of the left--hand side of the above
equation and substitution $u_t = - \varphi(u) u_x$ result
in an expression containing the first partial derivatives
of $\xi^{(t)},\xi^{(x)}$ and $\eta$ and also the factors
 $u_x$ and $u_x^2$. Since the resulting expressions must
equal zero for all values of $u_x$, the terms with different
powers of $u_x$ must be equated to zero separately. Having
grouped the terms by powers of $u_x$, one notices that
the terms with $u_x^2$ cancel each other. Equating the terms
without $u_x$ and the terms with $u_x$ to zero leads
to the following system of linear PDEs for the unknown
components of $X$:
\begin{eqnarray}
  \eta_t + \varphi(u) \eta_x &=& 0, \label{eq:det_system_1} \\
  \varphi(u) \xi^{(t)}_t - \xi^{(x)}_t - \varphi(u) \xi^{(x)}_x
 +\varphi^2(u) \xi^{(t)}_x + \varphi^{\prime}(u) \eta &=& 0.
                                    \label{eq:det_system_2}
\end{eqnarray}
Having solved the above system, one obtains the following
infinitesimal operator describing all o.p.g. admitted by
the nonlinear advection equation:
\begin{eqnarray}
 X = \xi^{(t)} \; \partial_t
   + ( A
      + \varphi(u)\xi^{(t)}
      + t \varphi^{\prime}(u) B) \; \partial_x
   + B \; \partial_u,
                                      \label{eq:advection_equation_groups}
\end{eqnarray}
where $\xi^{(t)}= \xi^{(t)}(t,x,u), A = A(u,x-\varphi(u)t)$ and
 $B = B(u,x-\varphi(u)t)$ are arbitrary smooth functions of the
indicated  arguments.

\subsection{Finite difference discretization and differential approximation}
A discretization of (\ref{eq:advection}) is performed on a stencil composed of
three (non--collinear) nodes of an unstructured grid -- $(t_0,x_0)$,
 $(t_1,x_1)$ and $(t_2, x_2)$. Denoting the values of the grid
 function at the nodes by $u_i$ and also $\varepsilon \tau_i = t_i - t_0$,
 $\varepsilon h_i = x_i - x_0$, ($i=1,2$), where $\varepsilon$
is a small parameter, the finite difference discretization
can be written as:
\begin{eqnarray}
   L_h(t_0,x_0,u_0,\varepsilon\tau_1,\varepsilon h_1,u_1,
                   \varepsilon\tau_2,\varepsilon h_2,u_2)
\equiv \frac{   h_2(u_1 - u_0) -    h_1(u_2 - u_0)}{\tau_1 h_2 - \tau_2 h_1}
+  \varphi(u)
       \frac{\tau_2(u_1 - u_0) - \tau_1(u_2 - u_0)}{\tau_2 h_1 - \tau_1 h_2}
   = 0. \label{eq:fd_approximation}
\end{eqnarray}
One can eliminate $u_1$ and $u_2$ from (\ref{eq:fd_approximation}) using
Taylor series expansions centered at the node $(t_0,x_0)$.
Omitting the index $0$ in the resulting expression, one
can write the result  as
\begin{eqnarray}
         L_0 + \varepsilon L_1 + o( \varepsilon ) = 0,
                                 \label{eq:FDA}
\end{eqnarray}
where
\begin{eqnarray*}
        L_0 &=& L(x,t,u,u_t,u_x) \equiv u_t + \varphi(u) u_x, \\
        L_1 &=& \frac{h_2 - \varphi(u) \tau_2}{2(\tau_1 h_2 - \tau_2 h_1)}
                (\tau_1^2 u_{tt} + 2 \tau_1 h_1 u_{tx} + h_1^2 u_{xx})
            - \frac{h_1 - \varphi(u) \tau_1}{2(\tau_1 h_2 - \tau_2 h_1)}
                (\tau_2^2 u_{tt} + 2 \tau_2 h_2 u_{tx} + h_2^2 u_{xx}).
\end{eqnarray*}
Equation (\ref{eq:FDA}) is called a {\em first differential
 approximation} \cite{SHOKIN:1} or an {\em equivalent differential equation}
\cite{KARNEY}, and it can be interpreted as a partial differential
equation with a small parameter which models the perturbations
of the original advection process due to the effects of finite difference
discretization.
Along with the small parameter $\varepsilon$,  equation (\ref{eq:FDA})
has the parameters $\tau_i$ and $h_i$ ($i=1,2$), which are not
necessarily small.

\subsection{Approximate Group Analysis
            of the First Differential Approximation}
Following \cite{BAIKOV:1,BAIKOV:2}, consider a small perturbation
of the infinitesimal operator $X = X_0 + \varepsilon X_1 + o(\varepsilon)$.
This perturbed operator corresponds to an {\em approximate} o.p.g. (a.o.p.g.)
 of transformations ${\bf z}^\prime = {\bf f}_0({\bf z},  a) + \varepsilon
 {\bf f}_1({\bf z},  a) + o(\varepsilon)$ by virtue of an ``approximate''
 Lie theorem \cite{BAIKOV:1}. The main problem of approximate group analysis
 is to find all a.o.p.g. of transformations such that the equation
 (\ref{eq:FDA}) remains invariant {\em up to $o(\varepsilon)$ terms}
 under these transformations. As in the case of ``exact'' group analysis,
 this problem can be solved through the approximate infinitesimal
 criterion of invariance \cite[Theorem 3, p.98]{BAIKOV:1} which, for
 this case, reduces to the solution of the problem
\begin{eqnarray}
  \left.
    \left(
           X_0^{(2)} + \varepsilon X_1^{(2)}
    \right)
    \left(
           L_0 + \varepsilon L_1
    \right)
  \right|_{L_0 + \varepsilon L_1 = o(\varepsilon)} = o(\varepsilon),
                          \label{eq:approximate_infinitesimal_invariance}
\end{eqnarray}
where
\begin{eqnarray*}
  X_i^{(2)}&=&  \xi_i^{(t)}      \partial_t
              + \xi_i^{(x)}      \partial_x
              + \eta_i           \partial_u
              + \zeta_i^{(t)}    \partial_{u_t}
              + \zeta_i^{(x)}    \partial_{u_x}      \\
           & &+ \zeta_i^{(tt)}   \partial_{u_{tt}}
              + \zeta_i^{(tx)}   \partial_{u_{tx}}
              + \zeta_i^{(xx)}   \partial_{u_{xx}}
              + \xi_i^{(\tau_1)} \partial_{\tau_1}
              + \xi_i^{(\tau_2)} \partial_{\tau_2}
              + \xi_i^{(h_1)}    \partial_{h_1}
              + \xi_i^{(h_2)}    \partial_{h_2},
\end{eqnarray*}
 $i=0,1$, and the terms $\zeta_i^{(tt)}, \zeta_i^{(tx)}$ and $\zeta_i^{(xx)}$
are obtained from the standard second prolongation formulae
\cite[Theorem 2.36]{OLVER}. As the above expression for $X_i^{(2)}$
shows, in addition to prolonging the action of these operators to
the second derivatives participating in (\ref{eq:FDA}), we must
also prolong it to the parameters $\tau_1, \tau_2, h_1$ and
 $h_2$ (the grid step sizes), because the latter are transformed
according to the transformations of the independent variables.
Consider, for instance, $\tau_1$. Since $\varepsilon \tau_1 = t_1 - t_0$,
we get:
\begin{eqnarray}
        \varepsilon \xi^{(\tau_1)}
\equiv  \varepsilon (\xi_0^{(\tau_1)}
                   + \varepsilon \xi_1^{(\tau_1)}
                   + o(\varepsilon))
= \xi^{(t)}(t_1,x_1,u_1)-\xi^{(t)}(t_0,x_0,u_0)
= \varepsilon ( \tau_1 D_t \xi_0^{(t)} + h_1 D_x \xi_0^{(t)})
  + o(\varepsilon).  \label{eq:expansion}
\end{eqnarray}
Equating the terms of the first order of $\varepsilon$ and performing
the same calculations for $\tau_2,h_1$ and $h_2$, one obtains:
\begin{eqnarray*}
  \begin{array}{ll}
    \xi_0^{(\tau_1)} = \tau_1 D_t \xi_0^{(t)} + h_1 D_x \xi_0^{(t)}, &
    \xi_0^{(\tau_2)} = \tau_2 D_t \xi_0^{(t)} + h_2 D_x \xi_0^{(t)}, \\
    \xi_0^{(h_1)}    = \tau_1 D_t \xi_0^{(x)} + h_1 D_x \xi_0^{(x)}, &
    \xi_0^{(h_2)}    = \tau_2 D_t \xi_0^{(x)} + h_2 D_x \xi_0^{(x)}.
  \end{array}
\end{eqnarray*}

Using expansions up to the second order in $\varepsilon$ in
(\ref{eq:expansion}) one can obtain the expressions for
 $\xi_1^{(\tau_1)}, \xi_1^{(\tau_2)}, \xi_1^{(h_1)}$ and
 $\xi_1^{(h_2)}$ in a similar way.

Substituting the above relations into
(\ref{eq:approximate_infinitesimal_invariance}) and using
(\ref{eq:FDA}) to eliminate
$u_t = - \varphi(u)u_x - \varepsilon L_1 + o(\varepsilon)$,
the left--hand side of (\ref{eq:approximate_infinitesimal_invariance})
can be rewritten as
\begin{eqnarray}
         (X_0 L_0 &+& \varepsilon (X_0 L_1 + X_1 L_0))
   =   \eta_{0t} + \varphi(u) \eta_{0x} +
        u_x( \varphi(u) \xi^{(t)}_{0t} - \xi^{(x)}_{0t}
           - \varphi(u) \xi^{(x)}_{0x}
            +\varphi^2(u) \xi^{(t)}_{0x} + \varphi^{\prime}(u) \eta_0 )
                                                           \nonumber\\
   & & + \varepsilon L_1 ( u_x \xi^{(x)}_{0u} - \varphi(u) u_x \xi^{(t)}_{0u}
                          -\eta_{0u}+\xi^{(t)}_{0t}+\varphi(u) \xi^{(t)}_{0x})
       + \varepsilon X_0 L_1 + \varepsilon X_1 L_0 + o(\varepsilon).
                                             \label{eq:decomposition}
\end{eqnarray}
Equating the leading terms in the above expression to zero, as
required by (\ref{eq:approximate_infinitesimal_invariance}), and
noticing that the resulting equation must hold for all $u_x$, one
obtains that the functions $\xi^{(t)}_0, \xi^{(x)}_0$ and
 $\eta_0$ must satisfy the system of equations
(\ref{eq:det_system_1})--(\ref{eq:det_system_2}), resulting from
 application of infinitesimal criterion of invariance
 to the original equation (\ref{eq:advection}).

Continuing decomposition by powers of $\varepsilon$ in
(\ref{eq:decomposition}), consider the equation obtained
by equating the  $O(\varepsilon)$ terms to zero.
First notice that one can eliminate $u_t$ from the resulting
equation using the substitution $u_t = -\varphi(u) u_x$, because
taking the term $\varepsilon L_1$ from (\ref{eq:FDA}) into
account contributes only to the terms $O(\varepsilon^2)$ which
are not considered here. After $u_t$ is eliminated , one
is left with a partial differential equation for the
unknown functions $\xi_1^{(t)},\xi_1^{(x)}$ and $\eta_1$
which depend on independent variables $t, x$ and $u$.
The equation also contains $\tau_i, h_i$, ($i=1,2$) and
 $u_x, u_{tt}, u_{tx}, u_{xx}$. Since (\ref{eq:FDA}) has already
been used to eliminate $u_t$, the resulting equation must
hold for all values of the parameters $\tau_i, h_i$, ($i=1,2$),
 $u_x, u_{tt}, u_{tx}, u_{xx}$. This allows one to decompose
the equation further. To this end, notice that among all
 $O(\varepsilon)$ terms in (\ref{eq:decomposition}),
only the term $\varepsilon X_1 L_0$ is independent of
 $\tau_i, h_i$, ($i=1,2$). Therefore, the equation
can be decomposed to obtain the following system:
\begin{eqnarray}
     X_1 L_0 &=& 0,    \label{eq:1st_correction}   \\
      L_1 ( u_x \xi^{(x)}_{0u} - \varphi(u) u_x \xi^{(t)}_{0u}
                         -\eta_{0u}+\xi^{(t)}_{0t}
                         +\varphi(u) \xi^{(t)}_{0x})
      + X_0 L_1 &=& 0, \label{eq:condition_on_X_0}
\end{eqnarray}
where $u_t$ is replaced by $-\varphi(u)u_x$. It follows from
equation
(\ref{eq:1st_correction}) that in order for the a.o.p.g.
 $X_0 + \varepsilon X_1 + o(\varepsilon)$ to leave  equation
(\ref{eq:FDA}) invariant up to $O(\varepsilon)$ terms, it is
necessary that the components of $X_1$ satisfy the same system
of equations (\ref{eq:det_system_1})--(\ref{eq:det_system_2})
as the system that an infinitesimal operator of an o.p.g.
admitted by  advection equation (\ref{eq:advection}) satisfies.
It was also established above that $X_0$ must be admitted
by the original equation (\ref{eq:advection}). Notice
that (\ref{eq:condition_on_X_0}) imposes additional
conditions on $X_0$. Knowledge of these conditions
allows one to select those of the o.p.g.
(\ref{eq:advection_equation_groups}) of the advection
equation which can be made a.o.p.g. of the first
differential approximation  at the price of adding
a small correction $\varepsilon X_1$ to the group
in order to account for the effects of discretization.
Following the terminology of \cite{BAIKOV:1}, these groups
are said to be {\em inherited} by a perturbed equation (\ref{eq:FDA}).
Notice that an infinitesimal operator $X^{(1)}$ with
all zero components satisfies (\ref{eq:1st_correction}).
Therefore, an operator which is inherited by (\ref{eq:FDA})
leaves this equation invariant up to $O(\varepsilon)$ terms
even without a small correction.
The purpose of the following is to derive this set of
conditions on $X_0$ and to describe all groups
inherited by (\ref{eq:FDA}).

Substituting the expression for $L_1$ into the left--hand side
of (\ref{eq:condition_on_X_0}) and multiplying both sides of
the resulting equation by $2 (\tau_1 h_2 - \tau_2 h_1)$, one
obtains an equation with zero on the right--hand side and
a number of terms on the left--hand side. Each of these terms
has a factor of the form $\tau_1^k \tau_2^l h_1^m h_2^n$,
where $k,l,m$ and $n$ can be $0,1$ or $2$, and the terms
with the same factors can be grouped together. Since
(\ref{eq:condition_on_X_0}) must hold for all values
of $\tau_1,\tau_2, h_1$ and $h_2$, each group of terms
must be equated to zero independently. This procedure gives
the following system of equations:
\begin{eqnarray*}
\zeta_0^{(tt)} + 2 u_{tt} D_t \xi_0^{(t)} + 2 u_{tx} D_t \xi_0^{(x)}
 -\varphi(u) u_x u_{tt} \xi_{0u}^{(t)} + u_x u_{tt} \xi_{0u}^{(x)}
 - u_{tt} \eta_{0u} &=& 0, \\
-\varphi(u)
  (\zeta_0^{(tt)} + 2 u_{tt} D_t \xi_0^{(t)} + 2 u_{tx} D_t \xi_0^{(x)})
  +\varphi(u) u_{tt} \xi_{0x}^{(x)} + u_{tt} \xi_{0t}^{(x)}
  -\varphi(u) u_x u_{tt} \xi_{0u}^{(x)} - u_{tt} \varphi^{\prime}(u)\eta_0&& \\
  +\varphi^2(u) u_x u_{tt} \xi_{0u}^{(t)} + \varphi(u) u_{tt} \eta_{0u}
  -\varphi(u) u_{tt} \xi_{0t}^{(t)} - \varphi^2(u) u_{tt} \xi_{0x}^{(t)}
  &=& 0, \\
\zeta_0^{(tx)} + u_{tx} D_t \xi_0^{(t)} + u_{tt} D_x \xi_0^{(t)}
  + u_{xx} D_t \xi_0^{(x)} + u_{tx} D_x \xi_0^{(x)}
  - \varphi(u) u_x u_{tx} \xi_{0u}^{(t)} + u_x u_{tx} \xi_{0u}^{(x)}
  - u_{tx} \eta_{0u} &=& 0, \\
-\varphi(u)
  (\zeta_0^{(tx)} +  u_{tx} D_t \xi_0^{(t)} +  u_{tt} D_x \xi_0^{(t)}
   + u_{xx} D_t \xi_0^{(x)} + u_{tx} D_x \xi_0^{(x)})
  +\varphi(u) u_{tx} \xi_{0x}^{(x)} + u_{tx} \xi_{0t}^{(x)} & & \\
  - \varphi(u) u_x u_{tx} \xi_{0u}^{(x)}
  - \varphi^{\prime}(u) u_{tx} \eta_0
  + \varphi^2(u) u_x u_{tx} \xi^{(t)}_{0u}
  + \varphi(u) u_{tx} \eta_{0u}
  - \varphi(u) u_{tx} \xi_{0t}^{(t)}
  - \varphi^2(u) u_{tx} \xi^{(t)}_{0x} &=& 0, \\
\zeta_0^{(xx)} + 2 u_{tx} D_x \xi_0^{(t)} + 2 u_{xx} D_x \xi_0^{(x)}
 -\varphi(u) u_x u_{xx} \xi_{0u}^{(t)} + u_x u_xx \xi_{0u}^{(x)}
 - u_{xx} \eta_{0u} &=& 0, \\
-\varphi(u)
  (\zeta_0^{(xx)} + 2 u_{tx} D_x \xi_0^{(t)} + 2 u_{xx} D_x \xi_0^{(x)})
  +\varphi(u) u_{xx} \xi_{0x}^{(x)} + u_{xx} \xi_{0t}^{(x)}
  -\varphi(u) u_x u_{xx} \xi_{0u}^{(x)} - u_{xx} \varphi^{\prime}(u)\eta_0&& \\
  +\varphi^2(u) u_x u_{xx} \xi_{0u}^{(t)} + \varphi(u) u_{xx} \eta_{0u}
  -\varphi(u) u_{xx} \xi_{0t}^{(t)} - \varphi^2(u) u_{xx} \xi_{0x}^{(t)}
  &=& 0.
\end{eqnarray*}
After the
expressions for $\zeta_0^{(tt)},\zeta_0^{(tx)}$ and $\zeta_0^{(xx)}$
and the total derivatives occuring in the above equations are
substituted, and $u_t$ is replaced by $-\varphi(u)u_x$, one finds
that in each equation the terms containing the second partial
derivatives of $u$ cancel each other.
Proceeding as before and decomposing
each of the above equations results in the
following set of necessary and sufficient conditions, under
which $X_0$ is inherited by (\ref{eq:FDA}):
\begin{eqnarray}
 \eta_{0tt} &=& 0, \label{eq:AI.1}\\
 \eta_{0tx} &=& 0, \label{eq:AIII.1}\\
 \eta_{0xx} &=& 0, \label{eq:AII.1}\\
 2 \varphi(u) \eta_{0tu} - \varphi(u) \xi_{0tt}^{(t)} + \xi_{0tt}^{(x)}&=&0,
                                              \label{eq:AI.2}               \\
              \eta_{0tu} - \xi_{0tx}^{(x)}
             -\varphi(u) \eta_{0xu} + \varphi(u) \xi_{0tx}^{(t)} &=& 0,
                                              \label{eq:AIII.2}              \\
 2 \eta_{0xu} - \xi_{0xx}^{(x)} + \varphi(u) \xi_{0xx}^{(t)}&=&0,
                                              \label{eq:AII.2}               \\
 2 \xi_{0tu}^{(x)} - 2 \varphi(u) \xi_{0tu}^{(t)}
                     + \varphi(u) \eta_{0uu}     &=& 0,
                                              \label{eq:AI.3}                \\
 - \varphi^2(u) \xi_{0xu}^{(t)} - \xi_{0tu}^{(x)} - \varphi(u) \eta_{0uu}
 + \varphi(u) \xi_{0tu}^{(t)} + \varphi(u) \xi_{0xu}^{(x)} &=& 0,
                                              \label{eq:AIII.3}              \\
 2 \varphi(u) \xi_{0xu}^{(t)} - 2 \xi_{0xu}^{(x)} + \eta_{0uu}  &=& 0,
                                              \label{eq:AII.3}               \\
 \xi_{0uu}^{(x)} - \varphi(u) \xi_{0uu}^{(t)} &=& 0.
                                              \label{eq:A*.4}
\end{eqnarray}

Since $\xi_0^{(t)},\xi_0^{(x)}$ and $\eta_0$ satisfy
(\ref{eq:det_system_1})--(\ref{eq:det_system_2}), one can write
 $\xi_0^{(t)}= \xi_0^{(t)}(t,x,u)$,
 $\xi_0^{(x)} = A(u,x-\varphi(u)t)
      + \varphi(u)\xi^{(t)}(t,x,u)
      + t\varphi^{\prime}(u) B(u,x-\varphi(u)t)$ and
 $\eta_0 = B(u,x-\varphi(u)t)$,
where $\xi^{(t)}, A$ and
 $B$ are arbitrary smooth functions of their arguments.
Introducing new independent variables $x_1=u, x_2=x-\varphi(u)t, x_3 = t$
and denoting the partial derivatives of the above arbitrary
functions by subindices (eg. $A_{12} = A_{x_1 x_2}$ etc),
system (\ref{eq:AI.1}) -- (\ref{eq:A*.4}) can be rewritten
as:
\begin{eqnarray}
 B_{22} &=& 0, \label{eq:B*.1} \\
   2\varphi  B_{12} + 4 \varphi^{\prime} B_{2} - \varphi  A_{22} &=& 0,
               \label{eq:BI.2} \\
 - 2\varphi  B_{12} - 2 \varphi^{\prime} B_{2} + \varphi  A_{22} &=& 0,
               \label{eq:BIII.2} \\
 A_{22} - 2 B_{12} &=& 0,
               \label{eq:BII.2}  \\
 - 2\varphi^{\prime}  A_2 - 2\varphi  A_{12}
 + 2 t \varphi  \varphi^{\prime}  A_{22}
 + 2   \varphi^{\prime}  \xi_{03}^{(t)}
 - 2 t \varphi  \varphi^{\prime}  \xi_{02}^{(t)}
 + 2   \varphi^{\prime\prime}  B                   & &  \nonumber\\
 + 2   \varphi^{\prime}  B_1
 - (4 t(\varphi^{\prime} )^2 + 3t\varphi  \varphi^{\prime\prime} ) B_2
 - 4 t \varphi  \varphi^{\prime}  B_{12} + \varphi  B_{11} &=& 0,
               \label{eq:BI.3} \\
 2   \varphi  A_{12} + \varphi^{\prime}  A_2
 - 2 t \varphi  \varphi^{\prime}  A_{22}
 - \varphi^{\prime}  \xi_{03}^{(t)}
 + 2 \varphi  \varphi^{\prime}  \xi_{02}^{(t)}
 - \varphi^{\prime\prime}  B                     & &  \nonumber\\
 - \varphi^{\prime}  B_1
 + (2 t(\varphi^{\prime} )^2 + 3 t\varphi  \varphi^{\prime\prime} ) B_2
 + 4 t \varphi  \varphi^{\prime}  B_{12}
 - \varphi  B_{11} &=& 0,
               \label{eq:BIII.3} \\
 2 A_{12}
 - 2 t \varphi^{\prime}  A_{22}
 + 2 \varphi^{\prime}  \xi_{02}^{(t)}
 + 3 t  \varphi^{\prime\prime}  B_2
 + 4 t \varphi^{\prime}  B_{12}
 - B_{11}     &=&  0,
               \label{eq:BII.3} \\
 A_{11}
 - 2 t \varphi^{\prime}  A_{12}
 - t \varphi^{\prime\prime}  A_{2}
 + t^2 (\varphi^{\prime} )^2 A_{22}
 + \varphi^{\prime\prime}  \xi_0^{(t)}
 + 2 \varphi^{\prime}  \xi_{01}^{(t)}
 - 2 t (\varphi^{\prime} )^2 \xi_{02}^{(t)}    & &  \nonumber\\
 + t \varphi^{\prime\prime\prime}  B
 + 2 t \varphi^{\prime\prime}  B_1
 - 3 t^2 \varphi^{\prime}  \varphi^{\prime\prime}  B_2
 + t \varphi^{\prime}  B_{11}
 - 2 t^2 (\varphi^{\prime} )^2 B_{12} &=& 0,
             \label{eq:B*.4}
\end{eqnarray}
It follows from ({\ref{eq:BI.2}}) and ({\ref{eq:BIII.2}}) that
 $\varphi^{\prime}B_2=0$. According to this, consider
two possibilities: $\varphi^{\prime}=0$ or $B_2=0$.

\subsubsection{Case $\varphi^{\prime}=0$ (linear advection)}
Denote $\varphi(u)= c \equiv const$. Using (\ref{eq:B*.1}),
write $B = B^{(0)}(x_1) + B^{(1)}(x_1) x_2$, where
$B^{(0)}(x_1)$ and $B^{(1)}(x_1)$ are unknown
functions.
Then (\ref{eq:BI.2}),
(\ref{eq:BII.2}), (\ref{eq:BIII.2}) give
 $A=(B^{(1)}(x_1))^{\prime} x_2^2 + g^{\prime}(x_1)x_2 + l(x_1)$
with $g(x_1)$ and $l(x_1)$ being unknown functions.
Substitution of the above expressions into the left--hand
side of (\ref{eq:BII.3})
results in an expression which is linear in $x_2$ with
its coefficients being functions of $x_1$. This expression
can be zero only if its coefficients are zero. This
results in two equations
 $(B^{(0)}(x_1) - 2 g(x_1))^{\prime\prime} = 0$ and
 $(B^{(1)}(x_1))^{\prime\prime} = 0$,
and therefore
 $B^{(0)}(x_1)= 2 g(x_1)+C x_1 + D $,
 $B^{(1)}(x_1)= E x_1 + F$ and
 $A = E x^2_2 + (g(x_1))^{\prime} x_2 + l(x_1)$
 where
 $C, D, E$ and $F$ are arbitrary constants.
Finally, (\ref{eq:B*.4}) implies that $A_{11} = 0$, and after substituting
the above expression for $A$, one obtains $(g(x_1))^{\prime\prime\prime}=0$
and $(l(x_1))^{\prime\prime} = 0$. Therefore
 $g(x_1) = M x^2_1 + N x_1 + P$ and
 $l(x_1) = K x_1 + L$, with arbitrary constants $K, L, M, N$
 and $P$.

Collecting the above yields the following expressions
for $A$ and $B$:
\begin{eqnarray*}
  A &=& E x_2^2 + (2 M + N) x_2 + K x_1 + L, \\
  B &=& 2 M x_1^2 + (2 N + C) x_1 + D + (E x_1 + F) x_2.
\end{eqnarray*}
Notice that, as follows from (\ref{eq:B*.1}) -- (\ref{eq:B*.4}),
 $\xi_{0}^{(t)}$ can still be an arbitrary function.

Since the expressions for $A$ and $B$ are linear
with respect to eight arbitrary constants, the set
of infinitesimal operators inherited by (\ref{eq:FDA})
is spanned by the following eight infinitesimal operators
\begin{eqnarray}
\begin{array}{ll}
  X_{(1)} = u \partial_u,       &  X_{(5)} = u \partial_x, \\
  X_{(2)} = \partial_u,         &  X_{(6)} = \partial_x,    \\
  X_{(3)} = (x-ct)^2\partial_x
          +u(x-ct)  \partial_u, &  X_{(7)} = u(x-ct)  \partial_x
                                             +(x-ct)^2\partial_u, \\
  X_{(4)} = (x-ct)\partial_u,   &  X_{(8)} =  (x-ct)\partial_x
                                              +2u  \partial_u
\end{array}
                                         \label{eq:varphi_1}
\end{eqnarray}
and an infinite--dimensional set of infinitesimal operators
\begin{eqnarray}
   X_{(\infty)} = \xi_{0}^{(t)} \partial_t + a \xi_{0}^{(t)} \partial_x,
                                          \label{eq:varphi_2}
\end{eqnarray}
where $\xi_{0}^{(t)} = \xi_{0}^{(t)}(t,x,u)$ is an arbitrary function.

\subsubsection{Case $\varphi^{\prime} \neq 0$, $B_2 = 0$}
In this case $B = B^{(0)}(x_1)$, where $B^{(0)}(x_1)$ is an unknown
function, and (\ref{eq:BII.2}) immediately gives $A_{22}= 0$, and
therefore $A = A^{(0)}(x_1) + x_2 A^{(1)}(x_1)$. Substituting
the expressions for $A$ and $B$ into (\ref{eq:BII.3}) gives
 $\xi^{(t)}_0 = x_2 ((B^{(0)})^{\prime\prime}-2 (A^{(1)})^{\prime})
 / (2 \varphi^{\prime}) + f(x_1, x_3)$,
 where $f(x_1,x_2)$ is an unknown function. After this expression
for $\xi^{(t)}_0$ is substituted into (\ref{eq:BI.3}), one
obtains
$
 -  \varphi^{\prime} A^{(1)} +  \varphi^{\prime} f_3
 +  \varphi^{\prime\prime} B^{(0)}
 +  \varphi^{\prime} (B^{(0)})^{\prime} = 0.
$
All terms in this equation, except the second one, depend
solely on $x_1$. As a consequence, we must have $f_3 = g(x_1)$
and $f=g(x_1) x_3 + l(x_1)$ where $g(x_1)$ and $l(x_1)$ are
unknown functions. Using this result, one can write
$
A^{(1)} = g + a B^{(0)} + (B^{(0)})^{\prime},
$
where $a = \varphi^{\prime\prime}/\varphi^{\prime}$.

Collecting together the above results, one obtains
\begin{eqnarray*}
  A &=& A^{(0)} + (g + a B^{(0)} + (B^{(0)})^{\prime}) x_2, \\
  B &=& B^{(0)}, \\
  \xi^{(t)}_0 &=& G(x_1) x_2 + g(x_1) x_3 + l(x_1),
\end{eqnarray*}
where $G(x_1) = -2 g -2 (a B^{(0)})^{\prime} - (B^{(0)})^{\prime\prime}$.
Proceeding with the above, substitute the resulting expressions
for $A, B$ and $\xi^{(t)}_0$ into (\ref{eq:BIII.3}) and ({\ref{eq:B*.4}}).
The first equation will be satisfied identically, while the second
will result in an equation which has zero on its right--hand side
and a linear expression  with respect to the independent
variables $x_2$ and $x_3$ with its coefficients being functions
of $x_1$ only, on its right--hand side. Equating these coefficients
to zero results in the following conditions on the unknown
functions $A^{(0)}, A^{(1)}, B^{(0)}, g$ and $l$:
\begin{eqnarray}
 -  \varphi^{\prime\prime} A^{(1)} + \varphi^{\prime\prime} g
 + 2\varphi^{\prime} g^{\prime}
 +  \varphi^{\prime\prime\prime} B^{(0)}
 + 2\varphi^{\prime\prime} (B^{(0)})^{\prime} &=& 0,
                               \label{eq:x_3} \\
   (A^{(1)})^{\prime\prime} + \varphi^{\prime\prime} G
   + 2 \varphi^{\prime} G^{\prime} &=& 0,
                               \label{eq:x_2} \\
   (A^{(0)})^{\prime\prime} + \varphi^{\prime\prime} l
 + 2 \varphi^{\prime} l^{\prime} &=& 0.
                               \label{eq:1}
\end{eqnarray}
Substitution of the expressions for $A^{(1)}$ and $G$
into (\ref{eq:x_3}) and (\ref{eq:x_2}) results in a system
of two differential equations with two unknown functions,
 $B^{(0)}$ and $g$. Integrating this pair of equations yields
\begin{eqnarray}
 a B^{(0)} + 2 g &=& C_1,          \label{eq:C_1}\\
 b^{\frac{1}{2}} B^{(0)} &=& C_2   \label{eq:C_2}
\end{eqnarray}
where $C_1$ and $C_2$ are arbitrary constants and
 $b = a^2/2 - a^{\prime}$.
Equation (\ref{eq:1}) can be integrated to find
 $l$ in terms of $A^{(0)}$:
\begin{eqnarray}
  l = C_3 (\varphi^{\prime})^{-\frac{1}{2}}
    - \frac{1}{2} (\varphi^{\prime})^{-\frac{1}{2}}
      \int (\varphi^{\prime})^{-\frac{1}{2}}
           (A^{(0)})^{\prime\prime} d x_1, \label{eq:l}
\end{eqnarray}
where $C_3$ is an arbitrary constant. In order to
formulate the final result, consider two cases:
 $b=0$ and $b \neq 0$.

\noindent\underline{\bf Case $b=0$.} \\
In this case $\varphi = K (L-u)^{-1} +M$, where $L$ and $M$
are arbitrary constants and  $K$ is a non--zero constant,
since in the present case $\varphi^{\prime} \neq 0$.
As follows from (\ref{eq:C_2}), $B^{(0)} =B^{(0)}(x_1)$
can be an arbitrary function and $C_2 = 0$.
Using (\ref{eq:C_1}) and (\ref{eq:l}), one finds that
the linear space of the infinitesimal operators
inherited by  (\ref{eq:FDA}) in this case is spanned
by the following infinitesimal operators:
\begin{eqnarray}
  X_{(1)} &=& t \;\partial_t + x \; \partial_x, \label{eq:b_1}\\
  X_{(2)} &=& (\varphi^{\prime})^{\frac{1}{2}} \; \partial_t +
              \varphi (\varphi^{\prime})^{\frac{1}{2}} \; \partial_x,
                                                \label{eq:b_2}\\
  X_{(3)} &=& \tilde{l} \; \partial_t + ( (x-\varphi t) A^{(0)}
                                      + \varphi \tilde{l} ) \; \partial_x,
                                                \label{eq:b_3}\\
  X_{(4)} &=& (-\frac{(aB^{(0)})^{\prime}+(B^{(0)})^{\prime\prime}}
                     {2 \varphi^{\prime}}(x-\varphi t)
               -\frac{1}{2} a B^{(0)}t) \; \partial_t  \nonumber \\
          & &
             +( \frac{1}{2} a B^{(0)}+(B^{(0)})^{\prime}
               -\frac{\varphi}
                     {2 \varphi^{\prime}}
                ((aB^{(0)})^{\prime}+(B^{(0)})^{\prime\prime})
                (x-\varphi t)
               -\frac{1}{2} a B^{(0)} t \varphi
               + t \varphi^{\prime} B^{(0)}
              ) \; \partial_x                          \nonumber \\
          & &
             + B^{(0)} \; \partial_u                   \label{eq:b_4}
\end{eqnarray}
where $\tilde{l}$ is related to $A^{(0)}$ by (\ref{eq:l}) with
 $C_3=0$. Unlike $X_{(1)}$ and
 $X_{(2)}$, the operators $X_{(3)}$ and $X_{(4)}$ each represent an infinite
 family of operators, parametrized by two arbitrary
 functions $B^{(0)}=B^{(0)}(u)$ and $A^{(0)}=A^{(0)}(u)$.

\noindent\underline{\bf Case $b \neq 0$.} \\
In this case, (\ref{eq:x_3}) and (\ref{eq:x_2}) immediately
give $B^{(0)} = C_2 b^{-\frac{1}{2}}$ and
 $g = \frac{1}{2} C_1 - \frac{1}{2} C_2 a b^{-\frac{1}{2}}$.
The functions $l$ and $A^{(0)}$, as in the previous case,
are related through (\ref{eq:l}). Using these expressions,
one can see that, as in the cases above,
all infinitesimal operators inherited by (\ref{eq:FDA})
form a linear space. For the given case, the basis in
this space can be chosen as follows:
\begin{eqnarray}
  X_{(1)} &=& t \;\partial_t + x \; \partial_x, \label{eq:general_1}\\
  X_{(2)} &=& (
               -\frac{b^{-\frac{5}{2}}}{8\varphi^{\prime}}
                  (4a^{\prime}b^2 - 2ab +3(b^\prime)^2-2bb^{\prime\prime})
                  (x-\varphi t)
               -\frac{1}{2} a b^{-\frac{1}{2}} t
              ) \; \partial_t                   \nonumber \\
          & &+(
               t \varphi^\prime b^{-\frac{1}{2}}
              -\frac{1}{2} t \varphi a b^{- \frac{1}{2}}
              + (
                  \frac{1}{2} b^{-\frac{3}{2}} (
                                                 a^{\prime\prime}
                                                -2 a a^\prime
                                                +\frac{a^3}{2}
                                               )
                 -\frac{\varphi b^{-\frac{5}{2}}}{8\varphi^{\prime}}
                  (4a^{\prime}b^2 - 2ab +3(b^\prime)^2-2bb^{\prime\prime})
                ) (x - \varphi t)
               ) \; \partial_x                  \nonumber  \\
          & & + b^{-\frac{1}{2}} \partial_u,    \label{eq:general_2} \\
  X_{(3)} &=& (\varphi^{\prime})^{\frac{1}{2}} \; \partial_t +
              \varphi (\varphi^{\prime})^{\frac{1}{2}} \; \partial_x,
                                                \label{eq:general_3} \\
  X_{(4)} &=& \tilde{l} \; \partial_t + ( (x-\varphi t) A^{(0)}
                                      + \varphi \tilde{l} ) \; \partial_x,
                                                \label{eq:general_4}
\end{eqnarray}
where  $\tilde{l}$ and  $A^{(0)}$
are related through (\ref{eq:l}) with $C_3=0$. Unlike the case
 $b=0$, only $X_{(4)}$ is an infinite family parametrized
 by an arbitrary function $A^{(0)}$.

\section{Results and Conclusion}
One parameter groups of transformations approximately
admitted by a first differential approximation corresponding
to a finite difference approximation of a nonlinear
advection equation have been studied.

It was demonstrated that the necessary condition for
such a group to be approximately admitted by a first
differential approximation is that this group is admitted
(exactly) by the original advection equation. Sufficient
conditions (\ref{eq:AI.1})--(\ref{eq:A*.4}), imposing
additional restrictions upon the
groups of the original equation, were also derived.

The description of a family of  transformation groups
admitted by a nonlinear advection equation
(\ref{eq:advection_equation_groups})
contains three arbitrary functions.
The requirement that the group must be inherited by
a first differential approximation (\ref{eq:FDA}) narrows
this arbitrariness in the following way.

For the most general case $\varphi^\prime \neq 0$, $b \neq 0$,
the family of transformation groups inherited by (\ref{eq:FDA})
is linear, and its basis may be chosen to consist of
three groups (\ref{eq:general_1})--(\ref{eq:general_3}) and
an infinite family of groups (\ref{eq:general_4}) parametrized
by an arbitrary function of $u$ only.

For the case $b=0$, the family of inherited groups expands and
can be described as a linear set spanned by two groups
(\ref{eq:b_1})--(\ref{eq:b_2}) and two infinite families
of groups (\ref{eq:b_3})--(\ref{eq:b_4}) each parametrized
by a different arbitrary function of $u$ only.

Finally, in the linear advection case when $\varphi^\prime = 0$,
the set of
inherited transformation groups has as its basis eight
groups (\ref{eq:varphi_1}) and one infinite family
of groups (\ref{eq:varphi_2}) parametrized by an arbitrary
function of $t, x$ and $u$.

\end{document}